\documentclass[aps,prb,showpacs,twocolumn]{revtex4}
\usepackage{amsfonts}
\usepackage{amssymb}
\usepackage{amsmath}
\usepackage{epsfig}
\usepackage{graphicx}

\begin{document}

\title{Generic Hubbard model description of semiconductor quantum dot spin qubits}
\author{Shuo Yang, Xin Wang, and S. Das Sarma}
\affiliation{Condensed Matter Theory Center, Department of Physics,
University of Maryland, College Park, MD 20742}

\begin{abstract}

We introduce a Hubbard model as the simple quantum generalization of the classical capacitance circuit model to study semiconductor quantum-dot spin qubits. We prove theoretically that our model is equivalent to the usual capacitance circuit model in the absence of quantum fluctuations. However, our model naturally includes quantum effects such as hopping and spin exchange. The parameters of the generalized Hubbard model can either be directly read off from the experimental plot of the stability diagram or be calculated from the microscopic theory, establishing a quantitative connection between the two. We show that, while the main topology of the charge stability diagram is determined by the ratio between inter-site and on-site Coulomb repulsion, fine details of the stability diagram reveal information about quantum effects. Extracting quantum information from experiments using our Hubbard model approach is simple, but would require the measurement resolution to increase by an order of magnitude.

\end{abstract}


\maketitle

\emph{Introduction-}
Electron spin qubits in coupled quantum-dot systems are among the most promising candidates for quantum information processing \cite{Loss.98,Levy.02,Koppens.06} because of their long coherence times \cite{Petta.05} and easy scalability.\cite{Taylor.05} Two physical effects dominate the electronic properties of quantum-dot systems: the Coulomb repulsion and the quantum fluctuations associated with electron hopping and spin exchange. \cite{Loss.98,Burkard.99,Koppens.06,Levy.02,Petta.05}

Due to Coulomb repulsion, the number of electrons on each dot can be controlled by gate voltages. \cite{Ciorga.00,Elzerman.03} This is mapped out by the charge stability diagram, \cite{Elzerman.03,SDexp} which illustrates the equilibrium charge configurations as a function of the two gate voltages on the dots. A classical capacitance circuit model neglecting all quantum effects, but including the Coulomb energy accurately, is almost universally used to model the charge stability diagram in quantum-dot systems.

Quantum effects are essential for encoding and manipulating quantum information in quantum-dot systems. In the simplest scheme, which makes use of single-electron spin-up (-down) states, the exchange interaction constitutes two-qubit operations.\cite{Loss.98} In the double-dot system, the singlet (triplet) states form the qubit, the level splitting of which is controlled by the exchange interaction.\cite{Taylor.05,Levy.02,Petta.05}

Since the charge stability diagram invariably serves as the experimental starting point for subsequent qubit manipulations, understanding it in depth is of great importance. Although the classical Coulomb effect is crucial in controlling the charge stability diagram, the quantum fluctuation effect should also manifest itself. A generic theory must be able to reconcile both.
Previously, the charge stability diagram has been studied using the classical capacitance model,\cite{RMP,Schroer.07} where the Coulomb interaction is parametrized by the electrostatic energy of effective capacitors. Although this method indeed produces the main experimental features,\cite{Elzerman.03,SDexp} the quantum effects are completely neglected in the capacitance model, which severely limits the applicability of the method in understanding quantum aspects of the charge stability diagram.

In this Rapid Communication, we introduce a Hubbard model, as the simplest quantum generalization of the classical capacitance model, including all possible quantum effects allowed by symmetry, and develop a quantum theory for the charge stability diagram. The Hubbard model has earlier been discussed in the literature \cite{DasSarma.9498,Gaudreau.06} in the context of quantum-dot physics, but a complete quantitative description of the charge stability diagram connecting to the capacitance circuit model and the microscopic confinement potential, as we provide here, has not been done before. We show that, in the absence of quantum effects, the generalized Hubbard model is equivalent to the capacitance model. However, while the quantum effects are intrinsically not allowed in the capacitance model, they can be straightforwardly accommodated in the generalized Hubbard model, making it suitable as a starting point of a more general theory including quantum fluctuations. We also note that some of the quantum effects were previously considered using a quantum mechanical two-level model,\cite{SDexp} which can be naturally derived from our more general Hamiltonian by projecting onto the single-electron subspace.

Microscopically, the quantum effects in the generalized Hubbard model originate from the overlap of the electron wave functions in a given confinement potential, which is treated using the configuration interaction method.\cite{Burkard.99,CI} For the bi-quadratic confinement potential,\cite{CI} we identify an important dimensionless parameter $\eta$ as the ratio between the height of the potential barrier and half the harmonic-oscillator energy-level spacing, which completely determines the parameters of the Hubbard model as well as the shape of the charge stability diagram. Since these parameters can be extracted from a comparison with the experimental results, the generalized Hubbard model provides a quantitative bridge between the experiment and the microscopic theory.

\emph{Mapping of capacitance model to Hubbard model-}
We consider a double quantum-dot system with each dot (labeled by $i=1,2$) capable of holding $N_i=0$, $1$ and $2$ electrons. Dots 1 and 2 are connected to the left and right gates with voltages $V_L$ and $V_R$, respectively.

We start with the classical capacitance model,\cite{RMP} in which
the two dots are connected with a capacitance $C_{m}$ and each dot is
connected to the nearest gate via capacitance $C_{L/R}$. The capacitance model
assumes exact integer number of electrons residing on each dot, which we
denote by $\left(N_{1},N_{2}\right)$.
The important quantity is the electrostatic energy as a function
of $N_{1}$ and $N_{2}$ \cite{RMP}:
\begin{equation}
\begin{split}
E\left(N_{1},N_{2}\right) = &\frac{1}{2}E_{C1}N_{1}^{2}+\frac{1}{2}E_{C2}N_{2}^{2}+E_{Cm}N_{1}N_{2}\\
&-E_{V1}N_{1}-E_{V2}N_{2}+E_{0}.
\end{split}
\label{Ecapacitance}
\end{equation}
Here, the charging energies are $E_{Cm}=\left|e\right|^{2}C_{m}/C_{\Sigma}^{2}$ and $E_{C1,2}=\left|e\right|^{2}\left(C_{R,L}+C_{m}\right)/C_{\Sigma}^{2}$.
Moreover, $E_{V1,2}=\left|e\right|\left(C_{L}C_{R}V_{L,R}+C_{L}C_{m}V_{L}+C_{R}C_{m}V_{R}\right)/C_{\Sigma}^{2}$, and $C_{\Sigma}^{2}=C_{L}C_{m}+C_{R}C_{m}+C_{L}C_{R}$.
Note that a precise value of the overall constant energy shift $E_0$ is unimportant for the study of charge stability diagram.

We map this model to the extended Hubbard model \cite{Imada.98} (without hopping), defined as
\begin{align}
H=\sum_{i=1,2}\left(-\mu_{i}n_{i}+U_{i}n_{i\uparrow}n_{i\downarrow}\right)+U_{12}n_{1}n_{2},\label{exthubmodel}
\end{align}
where $n_{i\sigma}=c_{i\sigma}^\dagger c_{i\sigma}$ is the number operator of electron with spin $\sigma$ on site $i$, $n_{i}=n_{i\uparrow}+n_{i\downarrow}$, $\mu_{i}$ and $U_{i}$ denote the chemical potential and Coulomb interaction on site $i$, respectively. $U_{12}$ denotes
inter-site Coulomb interaction between dots $1$ and $2$. Since $n_{1}$
and $n_{2}$ are good quantum numbers, the eigenenergy can be expressed
as
\begin{equation}
\!E'\left(n_{1},n_{2}\right)=\!\sum_{i=1,2}\!\left[-\mu_{i}n_{i}\!+\!\frac{U_{i}}{2}n_{i}\!\left(n_{i}-1\right)\right]\!+U_{12}n_{1}n_{2},
\label{EHubbard}
\end{equation}
where the on-site Coulomb repulsion is present only when the dot
is doubly occupied. $(\mu_{1}, \mu_{2})$ are linear combinations of
$\left(V_{L},V_{R}\right)$ plus a constant energy shift \cite{Gaudreau.06}:
$\mu_{1} = \left|e\right|\left(\alpha_{1}V_{L}+\beta_{1}V_{R}\right)+\gamma_{1}$, $\mu_{2} = \left|e\right|\left(\beta_{2}V_{L}+\alpha_{2}V_{R}\right)+\gamma_{2}$.
With this correspondence, comparing Eqs.~\eqref{Ecapacitance} and Eq.~\eqref{EHubbard},
a mapping is immediately found:
$U_{i}=E_{Ci}$, $U_{12}=E_{Cm}$, $\alpha_{1,2}=\left(C_{L}C_{R}+C_{L,R}C_{m}\right)/C_{\Sigma}^{2}$,
 $\gamma_{i}=-U_{i}/2$. In particular, when $U_{1}=U_{2}=U$, we
have
\begin{align}
1-\beta_{1,2}=\alpha_{1,2}=U / \left( U+U_{12}\right)\equiv\alpha.\label{capacalpha}
\end{align}

\begin{figure}
    \centering
    \includegraphics[width=7.8 cm, bb=30 679 225 778]{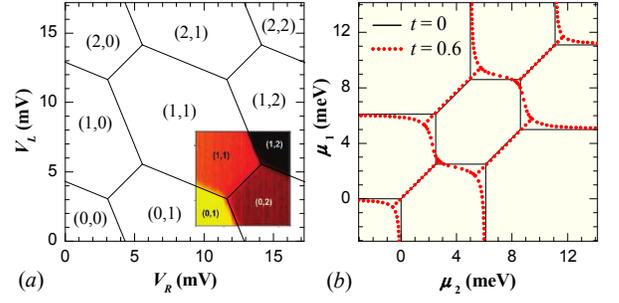}
    \caption{(Color online) (a) The charge stability diagram of the double dot system calculated using the extended Hubbard model [Eq.~\eqref{exthubmodel}], using parameters extracted from the experimental plot \cite{Petta.05} which has been overlaid in the lower right part of the figure. $U=6.1$ meV and $U_{12}=2.5$ meV. (b) Charge stability diagram calculated with the same $U$ and $U_{12}$ but with the hopping $t$ varied, plotted in the $\mu_1$-$\mu_2$ plane. $t=0$: black solid lines.  $t=0.6$ meV: red dotted lines. Note that the main effect of the quantum fluctuation is a rounding of the sharp triple point structure of the stability diagram.}
    \label{modelresult}
\end{figure}

Calculation of the charge stability diagram from Eq.~\eqref{exthubmodel} is straightforward. For given $\mu_1$ and $\mu_2$, one identifies the ground state [labeled by electron occupancy $(n_1, n_2)$] by finding the lowest eigenenergy. The geometry of the stability diagram is completely determined by $U$ and $U_{12}$, therefore one can read them off directly from the experimental plot.
 First, the length of the phase boundary between (1,1) and (0,2) is $\sqrt{2}U_{12}$, which defines $U_{12}$. Second, the slope of the phase boundary between (1,1) and (1,2) is $\alpha/(\alpha-1)$ and then $U$ is obtained from Eq.~\eqref{capacalpha}. Figure \ref{modelresult}(a) illustrates this idea, showing the charge stability diagram calculated from the extended Hubbard model, using parameters directly extracted from the experiment,\cite{Petta.05} which has been overlaid on the figure. The honeycomb shape is a direct consequence of a finite $U_{12}$, and therefore
the extended Hubbard model [Eq.~\eqref{exthubmodel}] is the minimal model that fits the experiment.

\emph{Generalized Hubbard model-}
Quantum fluctuations necessitate going beyond the extended Hubbard model. A general Hamiltonian for the double-dot system including up to two-body interactions can be written as
\begin{equation}
H=\sum_{kl}F_{kl}c_k^\dagger c_l+\sum_{hjkl}G_{hjkl}c_h^\dagger c_j^\dagger c_kc_l\label{genHam}
\end{equation}
where the subscripts include site and spin indices. In this paper, we focus on the case without magnetic field.
All possible terms that conserve the total particle number $N=n_{1}+n_{2}$ and the total spin $S_z$ can be classified into the following categories:
\begin{equation}
H=H_\mu+H_t+H_U+H_J\label{genHubmodel}
\end{equation}
where the chemical potential part is $H_\mu=\sum_{i\sigma}(-\mu_in_{i\sigma})$, and the hopping terms are $H_t=\sum_{\sigma}(-tc_{1\sigma}^\dagger c_{2\sigma}+\mathrm{H.c.})$. The Coulomb repulsion terms are $H_U=U_1n_{1\uparrow}n_{1\downarrow}+U_2n_{2\uparrow}n_{2\downarrow}+U_{12}(n_{1\uparrow}n_{2\downarrow}+n_{1\downarrow}n_{2\uparrow})
+(U_{12}-J_e)(n_{1\uparrow}n_{2\uparrow}+n_{1\downarrow}n_{2\downarrow})$, and $H_J=-J_ec_{1\downarrow}^\dagger c_{2\uparrow}^\dagger c_{2\downarrow}c_{1\uparrow}-J_pc_{2\uparrow}^\dagger c_{2\downarrow}^\dagger c_{1\uparrow}c_{1\downarrow}-\sum_{i\sigma}J_{t}n_{i\sigma}c_{1\overline{\sigma}}^\dagger c_{2\overline{\sigma}}+\mathrm{H.c.}$, including the spin-exchange ($J_e$),
pair-hopping ($J_p$), and occupation-modulated hopping terms ($J_t$) \cite{Imada.98}.
In the following, we assume $U_{1}=U_{2}=U$. Equation \eqref{genHubmodel} defines our proposed generalized Hubbard model.

When $H_t$ and $H_J$  are present, $n_1$ and $n_2$ are no longer good quantum numbers; therefore, the capacitance model ceases to be valid.
The Hamiltonian is, in general, a $16\times16$ matrix for the double-dot system but can be block diagonalized according to $\{N,S_z\}$. For states with different $N$, the phase boundaries are defined as before. For states within the same $N$ block, the ground state is a superposition of $(n_1,n_2)$ states and we label a phase according to the one that dominates the ground state.

Figure \ref{modelresult}(b) shows the calculated charge stability diagram with two different values of $t$ (with $J_e=J_p=J_t=0$), plotted on the $\mu_1$-$\mu_2$ plane. The $t=0$ data are the same as that shown in Fig.~\ref{modelresult}(a). Finite $t$ changes the curvature of various phase boundaries and thus smoothens the sharp corner of the tripl-point structure of the stability diagram. For instance, the phase separator between the (0,0) and (1,0)/(0,1) complex has been changed to hyperbola $\mu_1\mu_2=t^2$. Because of the smallness of $t$ and the insufficient resolution of the experiment in Fig.~\ref{modelresult}(a), the value of $t$ is not readily available from experiments at this stage, but $t$ can, in principle, be read off from the details of the $V_L$-$V_R$ diagrams.

The generalized Hubbard model has several tunable parameters. However, we have a caveat: the parameters can not be arbitrarily chosen, otherwise unphysical situations may occur. For example, if $U_{12}>U$, the (1,1) state is completely ruled out from the stability diagram.
Therefore, the  model must be backed up by a microscopic theory leading to a reasonable choice of the parameter set defining the Hubbard model.

\emph{Microscopic theory-}
In the microscopic theory,\cite{Burkard.99,CI} one typically solves the many-electron Schr\"odinger equation with the confinement potential $V\left(\boldsymbol{r}\right)$. The potential wells near the centers of the quantum dots are approximated as parabolic, so the low-lying excited eigenstates of an isolated single dot are of harmonic-oscillator type, i.e., the Fock-Darwin states. The Fock-Darwin state on dot $i$ can be written as $\left|\varphi_{i,n_{i},m_{i}}\left(\boldsymbol{r}_i\right)\right\rangle$, with the principal quantum number $n_i=0,1,\cdots$ and the azimuthal quantum number $m_{j}=-n_j,-n_j+2,
\cdots,n_j-2,n_j$, the meaning of which is analogous to that of atomic orbitals.

There are an infinite number of orbitals, but, for simplicity, we keep the lowest one denoted as $S$ orbital. The situation with multiple orbitals will be discussed elsewhere. The Fock-Darwin states for different dots are, in general, not orthogonal; therefore, we build a new set of orthogonal
basis by making the transformation
$\left(\left|\Psi_{1}\right\rangle \left|\Psi_{2}\right\rangle\right)^{\mathrm{T}}={\bf O}^{-1/2}\left(
\left|\varphi_{1}\right\rangle \left|\varphi_{2}\right\rangle\right)^{\mathrm{T}}$,
where ${\bf O}$ is the overlap matrix ($O_{kl}=\left\langle \varphi_{k}\right|\left.\varphi_{l}\right\rangle $) generated by the Fock-Darwin states in
the single particle subspace.
We recognize that the new basis $\left|\Psi_{l}\right\rangle$ actually corresponds
to $c_{l}^\dagger\left|0\right\rangle$. The coupling parameters in Eq.~\eqref{genHam} are $F_{kl}=\int d\boldsymbol{r}\Psi_{k}^{*}(\boldsymbol{r})h(\boldsymbol{r})\Psi_{l}(\boldsymbol{r})$, and $G_{hjkl}= \int d\boldsymbol{r}_{1} \int d\boldsymbol{r}_{2} \Psi_{h}^{*} (\boldsymbol{r}_{1}) \Psi_{j}^{*}(\boldsymbol{r}_{2})H_{C}\Psi_{k}(\boldsymbol{r}_{1})\Psi_{l}(\boldsymbol{r}_{2})$
where $h\left(\boldsymbol{r}\right)= \left[\boldsymbol{p}-e\boldsymbol{A}\left(\boldsymbol{r}\right)\right]^{2}$ $/(2m^{*})+V\left(\boldsymbol{r}\right)$ is the single-particle Hamiltonian, and $H_{C}=e^{2}/(4 \pi \varepsilon \varepsilon_{0} |\boldsymbol{r}_{1}-\boldsymbol{r}_{2}|)$ is the Coulomb interaction, with the effective mass of electrons
$m^{*}=0.067 m_{e}$ and relative dielectric constant $\varepsilon=13.1$ in GaAs quantum dot systems.

\begin{figure}[tbp]
\centering
\includegraphics[width=8 cm]{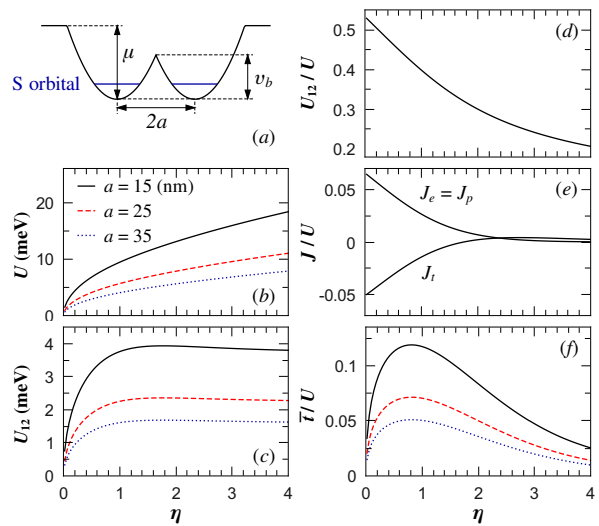}
\caption{(Color online) (a) \label{uvj} Schematic illustration for the bi-quadratic potential profile. $2a$ denotes the inter-dot distance, while $\upsilon_b$ represents the height of the central potential barrier when $\mu_1=\mu_2=\mu$. (b)-(f) Parameters in the generalized Hubbard model as functions of $\eta=m\omega_0a^2/\hbar$, calculated at $a=15$nm (black solid lines), 25nm (red dashed lines) and 35nm (blue dotted lines).
In panels (d) and (e), the calculated parameters do not change with $a$ so only the black solid lines are shown.
}
\end{figure}

In this calculation, we use a bi-quadratic potential shown in Fig.~\ref{uvj}(a) \cite{CI},
$V(x,y)={\rm Min} [W(-a,0)-\mu_{1}, W(a,0)-\mu_{2}, 0]$,
where the parabolic potential centered at $(x_{0},y_{0})$ is $W(x_{0},y_{0})=m \omega_{0}^{2} [(x-x_{0})^{2}+(y-y_{0})^{2}]/2$.
The important parameters are the inter-dot distance $2a$ and a dimensionless parameter $\eta=m\omega_0 a^2/\hbar$, denoting the ratio of the height of the central barrier ($\upsilon_{b}=m\omega_0^2a^2/2$) and
half the harmonic oscillator energy level spacing ($\hbar\omega_0/2$) when $\mu_{1}=\mu_{2}$.

Figure \ref{uvj} shows the microscopically calculated parameters of the generalized Hubbard model as functions of $\eta$ for three different values of $a$. We note that the $\eta<1$ case should be treated with caution: The central barrier is too low relative to the zero-point energy, effectively merging the two dots into one with very strong quantum fluctuations. Hereafter, we focus on the $\eta>1$ regime.
The on-site Coulomb interaction $U$ is relevant for the extent of the localization of electron wave functions, controlled by $\omega_0$, while the inter-site Coulomb interaction $U_{12}$ depends on the inter-dot distance $2a$. Figure \ref{uvj}(b) and \ref{uvj}(c) show these dependences: For a given $a$, increasing $\eta$ means that the electron wave functions become more localized, thus, $U$ increases correspondingly, while $U_{12}$ changes very slightly. If one fixes $\eta$ but increases $a$ (meaning $\omega_0$ must be decreased), both $U$ and $U_{12}$ decrease, as expected.

\begin{figure}[tbp]
\centering
\includegraphics[width=7 cm, bb=10 253 519 760]{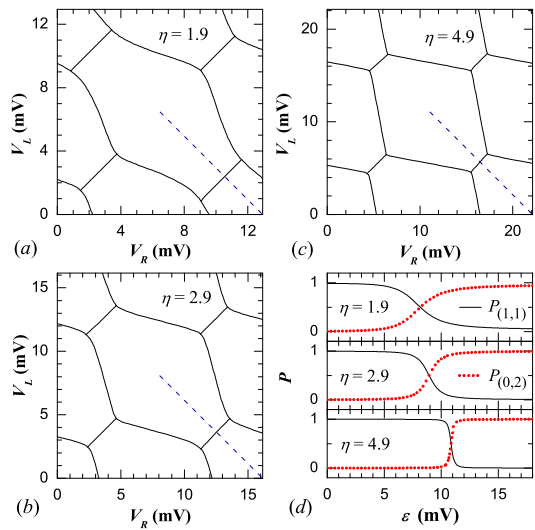}
\caption{(Color online) (a)-(c) Charge stability diagram calculated at three different values of $\eta$ with $a$ fixed as $33.8$ nm. (d) The probability of electron configurations $(1,1)$ and $(0,2)$ along the blue dashed lines on panels (a)-(c) as a function of the detuning energy $\varepsilon=V_R-V_L$. Black solid lines: the probability of (1,1) state; red dotted lines: the probability of (0,2) state.}
\label{varyeta}
\end{figure}

Unlike $U$ and $U_{12}$, which are constants over the entire stability diagram, $t$ slightly changes as the potential is varied. Thus we take an average value $\overline{t}$ over the range that we consider: $V_{L},V_{R} \in [0,(U+2 U_{12}+2 \hbar \omega_{0})/|e|]$ (see Fig.~\ref{varyeta}). $\overline{t}$ should decrease either as the dots become further apart from each other, or as the electron wave function in each dot becomes more localized with $\omega_0$ increasing. Figure \ref{uvj}(f) shows the ratio of $\overline{t}$ and $U$ as a function of $\eta$, and one can see that for $\eta>1$, $\overline{t}$ decreases as either $\eta$ or $a$ is increased. Figures \ref{uvj}(d) and \ref{uvj}(e) show $U_{12}$, $J_e$($=J_p$), and $J_t$ divided by $U$, respectively. The calculations have been carried out for all three values of $a$ but, remarkably, the results do not depend on any particular choice of $a$, leaving only a dependence on the dimensionless microscopic parameter $\eta$.

As mentioned above, $U_{12}/U$ plays a key role in determining the shape of the stability diagram, with $J_e$ and $J_t$ being an order of magnitude smaller than $U_{12}$. This can be seen in Fig.~\ref{varyeta}, which plots the stability diagrams for three different values of $\eta$. If the central potential barrier is low [Fig.~\ref{varyeta}(a)] the two dots effectively merge into one, and the curvature of the phase boundaries in the neighborhood of the triple point is very small. For an intermediate value of $\eta$ [Fig.~\ref{varyeta}(b)], the stability diagram has the honeycomb pattern. As $\eta$ further increases, the central barrier is high such that the overlap between the two dots is suppressed and the system behaves as two independent dots, leading to a tilted checker board pattern [Fig.~\ref{varyeta}(c)]. Therefore, our approach gives a quantitative description of the evolution of the charge stability diagram as the inter-dot quantum fluctuation is varied, which can be compared to experiments \cite{tunableexp} upon appropriately choosing parameters.\cite{Wang.10} As noted above, the boundaries between different configurations with the same $N$ are determined by the probability of configurations. As an example, Fig.~\ref{varyeta}(d) shows the probabilities of (1,1) and (0,2) states along the lines indicated as dashed lines in Figures \ref{varyeta}(a)-\ref{varyeta}(c), as functions of the detuning energy defined as $\varepsilon=V_R-V_L$. The crossing point defines the phase separator, and the width of the crossover encodes the quantum fluctuations. When $\eta$ is small, the inter-dot coupling is large: the states are substantially mixed and the crossover is rather smooth. If $\eta$ is large, the inter-dot coupling is small: the crossover is much sharper, as expected. Therefore, an alternative strategy of extracting quantum effects from experiments is to measure the width of the probability crossover near the phase separator rather than from the curvature of the phase boundary. Our technique can be applied quantitatively to experimental results with sufficient resolution.

\emph{Conclusion-} We present a generalized Hubbard model approach to the quantum-dot system to study the charge stability diagram including quantum fluctuations. We identify the capacitance model as the limiting case of the generalized Hubbard model with vanishing quantum fluctuations. We show that quantum effects have observable consequences on the charge stability diagram, which can be extracted using the Hubbard model. We establish a direct correspondence between the charge stability diagram and the microscopic double-dot confinement potential through the generalized Hubbard model, and show that the finer details of the stability diagram provide information about the underlying quantum effects.

\emph{Acknowledgement-} We thank X. Hu, J. P. Kestner, M. Cheng and A. J. Millis for helpful discussions. This work is supported by IARPA and LPS.

\end{document}